\documentclass[
  aip,
  cha,
  reprint,
  amsmath,amssymb,nofootinbib
]{revtex4-2}

\usepackage{graphicx}
\usepackage{bm}
\usepackage{dcolumn}
\usepackage{babel}
\usepackage{natbib}
\usepackage{color}
\usepackage{epstopdf}
\usepackage{amsmath} 
\usepackage{url}
\definecolor{darkGreen}{RGB}{0,110,0}
\definecolor{darkBlue}{RGB}{0,0,130}
\usepackage[colorlinks=true, urlcolor=darkBlue,citecolor=darkGreen,linkcolor=darkBlue]{hyperref}
\usepackage{amssymb}
\usepackage{comment}
\usepackage{physics}
\usepackage{float}
\usepackage{balance}

\usepackage[normalem]{ulem}

\begin{document}

\title{Forecasting Low-Dimensional Navier-Stokes equations via Multi‑Dimensional Hybrid Quantum Reservoir Computing}
\author{L. Salatino}
\affiliation{ICAR-CNR, Consiglio Nazionale delle Ricerche, Italy}

\author{L. Mariani}
\affiliation{ICAR-CNR, Consiglio Nazionale delle Ricerche, Italy}

\author{A. Giordano}
\affiliation{ICAR-CNR, Consiglio Nazionale delle Ricerche, Italy}

\author{F. D'Amore}
\affiliation{ICAR-CNR, Consiglio Nazionale delle Ricerche, Italy}

\author{C. Mastroianni}
\affiliation{ICAR-CNR, Consiglio Nazionale delle Ricerche, Italy}

\author{L. Pontieri}
\affiliation{ICAR-CNR, Consiglio Nazionale delle Ricerche, Italy}

\author{A. Vinci}
\affiliation{ICAR-CNR, Consiglio Nazionale delle Ricerche, Italy}

\author{C. Gencarelli}
\affiliation{IGAG-CNR, Consiglio Nazionale delle Ricerche, Italy}

\author{L. Primavera}
\affiliation{Dipartimento di Fisica, Università della Calabria, Via P. Bucci Arcavacata di Rende (CS), Italy}

\author{F. Plastina}
\affiliation{Dipartimento di Fisica, Università della Calabria, Via P. Bucci Arcavacata di Rende (CS), Italy}
\affiliation{INFN, Gruppo Collegato di Cosenza, Arcavacata di Rende (CS), Italy}

\author{J. Settino}
\affiliation{Dipartimento di Fisica, Università della Calabria, Via P. Bucci Arcavacata di Rende (CS), Italy}
\affiliation{INFN, Gruppo Collegato di Cosenza, Arcavacata di Rende (CS), Italy}
\email{jacopo.settino@unical.it}

\author{F. Carbone}
\affiliation{IIA-CNR, Consiglio Nazionale delle Ricerche, Italy}

\date{\today}
             
\begin{abstract}
\noindent{\textit{Corresponding author:} J. S. - } \href{mailto:jacopo.settino@unical.it}{jacopo.settino@unical.it}\\
The prediction of complex dynamics remains an open problem across many domains of physics, where nonlinearities and multiscale interactions severely limit the reliability of conventional forecasting methods. Quantum Reservoir Computing has emerged as a promising paradigm for information processing by exploiting the nonlinear dynamical evolution of quantum systems to generate expressive feature representations of input signals.
Here, we introduce a hybrid quantum-classical reservoir architecture capable of handling multivariate time series through quantum evolution combined with classical memory enhancement. Our model employs a five-qubit transverse-field Ising Hamiltonian with input-modulated dynamics and temporal multiplexing. We apply this framework to two paradigmatic models of chaotic behavior in fluid dynamics, where multiscale dynamics and nonlinearities play a dominant role: a low-dimensional truncation of the two-dimensional Navier-Stokes equations and the Lorenz-63 system.
By systematically scanning the quantum system's parameter space, we identify regions that maximize forecasting performance. The observed robustness and reliable performances for both dynamical systems suggest that this hybrid quantum approach offers a flexible platform for modeling complex nonlinear time series.

\end{abstract}

\maketitle
\begin{quotation}
Forecasting chaotic systems is fundamentally challenging because small uncertainties can grow rapidly, limiting predictive accuracy in nonlinear phenomena ranging from turbulence to climate dynamics. Here, we show that a hybrid quantum–classical reservoir can improve the prediction of such complex behavior by combining the expressive dynamics of a quantum system with a classical memory layer. We assess the method on two canonical benchmarks of chaos, a low-dimensional model of turbulence derived from the Navier–Stokes equations and the Lorenz–63 system, and show that temporally multiplexed measurements of the quantum dynamics substantially enhance forecasting performance. We also identify parameter regimes that remain effective across both benchmarks, suggesting a degree of universality in the proposed approach. In addition, the hybrid architecture is well-suited to quantum hardware with a small number of qubits,  because it naturally enables the parallel encoding of multiple input signals. Together, these results highlight hybrid quantum reservoir computing as a promising strategy for modeling and predicting nonlinear dynamical systems.    
\end{quotation}

\section{Introduction}
Accurately forecasting the behavior of chaotic systems remains a fundamental challenge across a wide range of disciplines, from fluid dynamics and meteorology to astrophysics and finance \cite{boffetta2002predictability, wu2024predicting}. These systems exhibit sensitive dependence on initial conditions and nonlinear interactions across multiple scales\cite{Lorenz1963, eckmann1985ergodic}, which limit the reliability of traditional prediction methods. In recent years, Reservoir Computing (RC) has emerged as an efficient and lightweight framework for modeling and predicting such complex dynamics~\cite{Jaeger2001, Maass2002, Pathak2018}. Inspired by recurrent neural networks, RC employs a fixed, high-dimensional dynamical system—the reservoir—that transforms time-dependent inputs into a rich feature space. A simple linear readout is then trained to extract relevant patterns from the reservoir's response, making RC particularly well-suited for time series prediction and control tasks~\cite{maass2002real,jaeger2004harnessing}.

The advent of quantum computing \cite{Shor1994, Grover1996, Nielsen2010, Montanaro2016, Preskill2018, Mastroianni2023,mariani2024simulation,salatino2023dissipative, Mastroianni2024, Consiglio2024,damore2025assessing}
has opened new perspectives for the development of RC schemes. Quantum Reservoir Computing (QRC) exploits the dynamical evolution of quantum systems to construct quantum reservoirs capable of processing information with high expressiveness. The initial concept of QRC, introduced by Fujii and Nakajima \cite{fujii2017quantum}, employs the space of quantum density matrices to store and process information by encoding input data into the state of a single qubit within a fully connected qubit network. Since the introduction of the QRC paradigm, a broad range of physical platforms has been investigated, each offering unique advantages depending on the context~\cite{ghosh2021quantum, zhu2025practical}. Among these, spin chains have attracted particular interest and have been widely applied to tasks such as time series forecasting and Extreme Learning Machines~\cite{sannia2024quantum, martinez2021physical, martinez2023cognitive, gotting2023physreva, kobayashi2024physreve, innocenti2023commphys, monaco2024machine, vetrano2024arxiv, palacios2024arxiv, monzani2024arxiv, ivaki2024arxiv,Fujii2017,DeLorenzis2025,settino2025topology,romeo2026extreme}. Other explored platforms include fermionic and bosonic systems, which have been proposed as alternative reservoirs~\cite{ghosh2019npjqi, tran2023physrevresearch, llodra2023aqt,romeo2025probing}, as well as quantum oscillators~\cite{govia2021physrevresearch}. 
Beyond these platforms, considerable effort has recently been devoted to identifying suitable hardware platforms and efficient information encoding strategies for quantum reservoir computing. Scalable photonic reservoirs have been proposed
for real-time quantum information processing and machine
learning applications~\cite{Garcia2022}, while experimental
control of memory properties has recently been demonstrated
in continuous-variable optical quantum reservoirs
\cite{Paparelle2026}. In parallel, QRC has also been
successfully applied to quantum-information tasks such as
photonic entanglement witnessing~\cite{Zia2025}. More recently, Rydberg atom arrays have been considered for
their strong and tunable dipole--dipole interactions,
offering a flexible platform for implementing complex
dynamics~\cite{bravo2022prxquantum,kornjaca2024arxiv}.
Gate-based quantum platforms have also been investigated
for QRC implementations due to their universality and
fine control capabilities
\cite{domingo2023arxiv,domingo2023scirep,yasuda2023arxiv,
wudarski2024arxiv,kubota2023physrevresearch,
fuchs2024arxiv,kobayashi2024physreve}. Several studies
have demonstrated the potential of QRC to outperform
classical reservoirs in tasks such as time-series
prediction, signal classification, and system
identification, especially in low-data or noisy
regimes~\cite{Govia2021,Kutvonen2020}.

While the exponential scaling of Hilbert-space dimension is often regarded as a key resource for QRC, practical implementations on near-term devices typically involve relatively small quantum systems. In this regime, the motivation for using quantum reservoirs is not necessarily associated with large-scale quantum computation, but rather with the ability of quantum many-body dynamics to generate nonlinear, high-dimensional embeddings of input signals via coherent evolution and multi-observable measurements.

In this work, we introduce a hybrid quantum-classical reservoir computing architecture designed to handle multivariate time series. The model allows for the simultaneous injection of multidimensional input data into the quantum reservoir, enabling efficient processing of complex high-dimensional dynamical signals. Memory is incorporated through classical post-processing of quantum measurements, following the scheme originally introduced in~\cite{settino2024memory}. 

An important aspect of this hybrid architecture is that temporal information is retained through the classical recurrent layer rather than through the intrinsic persistence of the quantum state itself. 

This choice avoids relying on the preservation of quantum-state memory across successive inputs and reduces the impact of measurement-induced state collapse. In this sense, the present approach differs from the original QRC framework introduced by Fujii and Nakajima~\cite{fujii2017quantum}, where the reservoir state itself carries temporal information, and measurements may directly perturb the ongoing quantum dynamics. Several recent works have addressed this issue through alternative strategies. Weak-measurement protocols have been proposed to continuously extract information while minimizing measurement back-action and preserving the memory stored in the quantum reservoir~\cite{Gebhart2023}. Other approaches employ restricted measurement schemes, where only a subset of the reservoir observables is monitored in order to reduce the disturbance induced by the measurement process~\cite{Kutvonen2022}. More recently, feedback-driven reservoir architectures have been introduced, where measurement outcomes are fed back into the quantum dynamics to recover fading-memory properties despite the use of projective measurements~\cite{Bausch2024,Sakurai2024}. In contrast, the present protocol follows a different strategy: the quantum subsystem is primarily used as a nonlinear feature generator, while temporal memory is delegated to a classical recurrent structure. This separation of roles naturally mitigates measurement-back-action issues and avoids the need for explicitly preserving memory within the quantum state.

The resulting architecture can be viewed as a Quantum Extreme Learning Machine equipped with a continuously tunable ~\cite{settino2024memory} or discrete~\cite{Mccaul2025minimal} classical memory. Its capability is further enhanced by incorporating a temporal multiplexing mechanism~\cite{Multiplexing} to increase the information content extracted from quantum evolutions at different timescales. The proposed hybrid QRC model is applied to a five-mode Galerkin truncation of the two-dimensional Navier-Stokes equations~\cite{Carbone2021, Boffetta2012}, providing a physically grounded and challenging setting for low-dimensional turbulence forecasting. For comparison, we also evaluate the model on the Lorenz-63 system~\cite{Lorenz1963}, a well-established benchmark for deterministic chaos. Moreover, we explore the dependence of prediction performance on the Hamiltonian parameters and evolution times, and we find that the optimal region coincides for both benchmarks, suggesting a potential generality of the proposed scheme. 
Compared with the lower-dimensional benchmarks previously considered in related hybrid-QRC studies, the Navier--Stokes and Lorenz-63 systems exhibit stronger nonlinear couplings, increased sensitivity to initial conditions, and more demanding long-term forecasting behavior. This increased dynamical complexity provides a more stringent benchmark for evaluating the robustness and predictive capability of the proposed architecture.

The article is structured as follows: in Sec.~\ref{sec:Method}, we introduce our hybrid‑QRC model for multivariate data and implement the temporal multiplexing scheme; In Sec.~\ref{sec:III}, we apply the algorithm to a five‑mode Galerkin truncation of the two‑dimensional Navier–Stokes equations and to the Lorenz-63 model, identifying the regions in the Hamiltonian parameter space and the evolution times that yield the highest predictive performance; Sec.~\ref{sec:conclusions} summarizes our conclusions. 

\newpage

\begin{figure}[h!btp]
    \centering
    \includegraphics[width=1\linewidth]{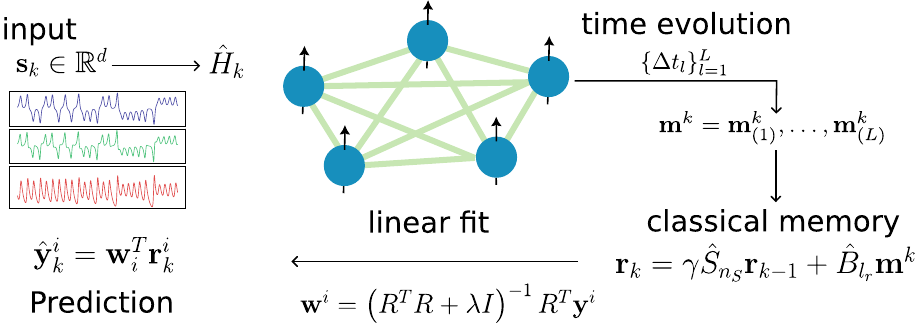}
   \caption{\small{\textbf{Schematic representation of the proposed hybrid QRC method.} At each step \( k \), the $d$-dimensional input $\mathbf{s}_k$ is encoded into the quantum reservoir by modulating some parameters of the Hamiltonian \( H_k \), which generates the dynamics of a quantum spin system. The system evolves for $L$ different time intervals \(\Delta t_l\), giving rise to the quantum states \(\lvert \psi_k(\Delta t_l) \rangle\) which depend on the input. After the time evolution, the output is obtained by concatenating the $L$ sets of measurements \(\mathbf{m}_{(l)}^{k}\), which carry information about the input data. These measurements are further processed classically to construct a reservoir state \(\mathbf{r}_k\) that retains the memory of previous inputs. A Ridge linear regression is then applied to the reservoir state to produce the predicted output.}}
    \label{fig:scheme}
\end{figure}

\section{Method}
\label{sec:Method}

We present a QRC model, in which the input was embedded into the system through modulation of a Hamiltonian parameter. Our model uses a Hamiltonian similar to the one proposed by Fujii and Nakajima~\cite{Fujii2017}, namely a transverse-field Ising model with an additional input-dependent local longitudinal magnetic field:
\begin{equation}
\hat{H}_k = \sum_{\substack{i,j=1 \\ i<j}}^N J_{ij} \hat{\sigma}_i^x \hat{\sigma}_j^x + h \sum_{i=1}^N \hat{\sigma}_i^z + \sum_{i=1}^N h_k^{i} \hat{\sigma}_i^x,
\end{equation}
where $J_{ij}$ are the coupling constants, $\hat\sigma^x,\hat\sigma^z$ are Pauli matrices, $h$ is a static transverse magnetic field,  $h_k^{i} = \sum_j \beta_j C_{ij} s_j^{k}$ is an input-dependent vector that encodes the input $\mathbf{s}_k \in \mathbb{R}^d$ at time step $k$ into the quantum system, through a constant $\beta_j$ and a matrix $\hat{C}\in \mathbb{R}^{N \times d}$. 
Given an initial pure state $|\psi_0\rangle= |0\rangle^{\otimes N}$, the quantum system evolves under the input-dependent Hamiltonian $\hat{H}_k$ for different evolution times $\{\Delta t_l\}_{l=1}^L$, leading to a set of $L$ quantum states:
\begin{align}
\ket{\psi_k(\Delta t_l)} &= e^{-i \hat{H}_k \Delta t_l} |\psi_0\rangle.
\end{align}

Hamiltonian-based input encoding has been widely employed in quantum reservoir computing and related quantum machine learning architectures, where external inputs modulate either local fields or interaction terms of the system Hamiltonian~\cite{fujii2017quantum, innocenti2023commphys, monaco2024machine}. In the present work, multidimensional inputs are simultaneously encoded through input-dependent longitudinal fields acting on different qubits.

After each evolution, a set of observables is measured, producing \(L\) measurement vectors $\mathbf{m}_{(l)}^{k}$. The final measurement vector  $\mathbf{m}^{k}$ is constructed by concatenating all of the \(L\) measurements:
\begin{equation}
\label{eq:measvec}
\mathbf{m}^{k}=\mathbf{m}_{(1)}^k, \space\dots\space ,\mathbf{m}_{(L)}^k
\end{equation}
The set of measured observables consists of:
\begin{equation}
\mathbf{m}_{(l)}^k=\left\{\left\langle\hat\sigma_{i}^{\alpha}\right\rangle\right\}_{i=1, \ldots, N}^{\alpha \in\{x, y, z\}} \cup\left\{\left\langle\hat\sigma_{i}^{\alpha} \hat\sigma_{j}^{\alpha}\right\rangle\right\}_{i<j}^{\alpha \in x, y, z},
\end{equation}
that are the expected values of the three spin components and the two-qubit Pauli correlations, along each axis and for every possible choice of qubit(s).
This method implements a \textit{temporal multiplexing} scheme that enhances the encoding of the $k$-th input by probing the quantum dynamics at different time scales. As shown in Fig.~\ref{fig:scheme}, this results in a richer and more informative measurement vector, effectively increasing the expressive power of the reservoir~\cite{steinegger2025predicting}.
For a system of $N$ qubits, the total number of measured observables for a single evolution time is $N_{obs} = 3N + 3\frac{N(N-1)}{2}$, corresponding respectively to all single-qubit Pauli expectation values and all pairwise Pauli correlations along the three spatial directions. (For $N$=$5$, this gives $N_{obs}$=$45)$. When temporal multiplexing with $L$ evolution times is employed, the measurement vectors obtained at different times are concatenated, leading to a total feature-space dimension $N_{feat} = L \cdot N_{obs}$. In the present work, we used $L=2$, resulting in $N_{feat}=90$.

To retain memory of previous inputs despite the time-local Hamiltonian, we construct a classical reservoir state inspired by the architecture of classical recurrent neural networks. The reservoir state \( \mathbf{r}_k \in \mathbb{R}^{l_r} \) evolves according to:
\begin{equation}
\mathbf{r}_k = \gamma \hat{S}_{n_S} \mathbf{r}_{k-1}+(1-\gamma)\hat{B}_{l_r}\mathbf{m}^{k},
\end{equation}
where \( \gamma \in [0,1] \) is a memory retention parameter, \( \hat{S}_{n_S} \) is a cyclic permutation operator that acts as \( [\hat{S}_{n_S}]_{ij} = \delta_{(i+n_S) \bmod l_r, j} \), where $\delta$ denotes the Kronecker delta, shifting the vector elements by $n_S$ steps forward, if $n_S$ is positive, or backward, if $n_S$ is negative, and \( \hat{B}_{l_r} \) is a linear operator that maps the concatenated measurement vector \( \mathbf{m}^{k} \) into a vector of length \( l_r \) by interleaving zeros, equally distributed, between its entries \cite{settino2024memory}. During the training phase, the reservoir states generated from \( N_{\mathrm{tr}} \) input steps are collected, and a linear readout is trained. The predicted values \( \hat{\mathbf{y}}_k^{i} \) at step \( k \) are given by:
\begin{equation}
\hat{\mathbf{y}}_k^{i} = \mathbf{w}^{T}_i \mathbf{r}_k^{i}, \quad \text{with} \quad \mathbf{w}^{i} = \left( R^T R + \lambda I \right)^{-1} R^T \mathbf{y}^{i},
\end{equation}
where \( R \) is the matrix of stacked reservoir vectors, \( \mathbf{y}^i \) is the vector of target values and \( \lambda \) is the Ridge regularization parameter \cite{Ridge1970}. Here, the index $i$ runs over the dataset. To identify the optimal combination of the quantum system and reservoir parameters that maximizes prediction performance, we conducted an extensive grid search over the relevant hyperparameters. Specifically, we scanned across values of the reservoir memory parameter \( \gamma \), the Hamiltonian coupling strength \( J \) (with the $J_{ij}$ randomly generated from a uniform distribution in $[-J, J]$), the transverse magnetic field \( h \), and the two evolution times $\Delta t _1$ and $\Delta t _2$, having fixed $L=2$. The optimal configuration was selected based on the highest average Valid Prediction Time, defined in Sec.~\ref{sec:NSRes} and obtained across multiple realizations of $\hat{H}_k$.  In the present work, a system of \( N = 5 \) qubits has been used. The dimension of the classical reservoir state was set to $l_r = 90$ and the memory parameter to $\gamma = 0.25$. A washout period of $N_{wo}$ = 200 time steps was
employed to remove the influence of the initial reservoir
state. 
In the following discussion, for each fixed $i=1\dots d$, the sequence  \(s^i_k\) is normalized to the range $[0,1]$, $\beta_j=1$ and $C_{ij}= \delta_{ij} $ for $j\leq d$, $0$ otherwise.  This choice consequently sets the units of energy and time.  
During the training phase, the target output corresponds to the next time step of the multivariate time series, i.e., the model is trained to learn the one-step forward prediction. In the testing phase, the model operates in an autonomous regime: the predicted output at time step $k$ is fed back as input for step $k+1$. This recursive procedure allows us to evaluate the long-term predictive capability of the reservoir.

\section{Forecasting of Chaotic Dynamics: {A low dimensional truncation of the 2D Navier-Stokes equations and Lorenz-63 system }}
\label{sec:III}

In this section, we evaluate the performance of our hybrid QRC model by applying it to a prediction task based on a low dimensional truncation of the two-dimensional Navier--Stokes (NS) equations (Eq.~\ref{eq:dynsist1}) for two distinct values of the forcing \( {F} \), corresponding to physical scenarios with different degrees of dynamical complexity and to the Lorenz-63 system (Eq.~\ref{eq:lorenz63}), a well-established benchmark for chaotic dynamics.

\subsection{Model}
\subsubsection{Navier-Stokes}

The non-dimensional Navier-Stokes (NS) system describes incompressible fluid dynamics in a two-dimensional periodic domain $0 \le (x,y) \le L \equiv 2\pi$. It is obtained from the projection of the momentum equations (Eq.~\ref{eq:mom1}) onto the coordinate axes and the continuity equation (Eq.~\ref{eq:cont1}). To account for the inverse energy cascade typical of 2D turbulence, a Rayleigh-type friction term $-\lambda_R \mathbf{u}$ is included, modeling unresolved dissipative effects such as bottom friction or sub-grid scale contributions. The governing equations read:
\begin{align}
\frac{\partial \mathbf{u}}{\partial t} +(\mathbf{u}\cdot\nabla)\mathbf{u} &= -\nabla P + \textnormal{Re}^{-1}\nabla^2\mathbf{u} -\lambda_R \mathbf{u} + \mathbf{F}\label{eq:mom1}\\
\nabla\cdot\mathbf{u} &= 0 \label{eq:cont1},
\end{align}

Here, $\mathbf{u} = [u_x, u_y]$ is the velocity field, $P$ is the normalized pressure, $\textnormal{Re}$ is the Reynolds number, and $\mathbf{F}$ is an external forcing. The derivation of the reduced dynamical representation follows standard procedures in Galerkin truncation of incompressible flows and is not repeated here for brevity (see Appendix~\ref{app:NS5} for technical details).

After truncation and restriction to the real invariant subspace, the dynamics reduce to a five-mode Lorenz-like system $\mathbb{L}_5(u)$, which captures the essential nonlinear interactions responsible for energy transfer among modes.

The resulting truncated Navier-Stokes model is given by:
\begin{equation}
\label{eq:dynsist1}
\begin{aligned}
\dot{u}_1 &= 3u_3 u_2 - u_1 \\
\dot{u}_2 &= -4u_3u_1 + 4u_5^2 - 2 u_2 \\
\dot{u}_3 &= u_1u_2 - 5u_3 \\
\dot{u}_4 &= 7u_5u_2 - 5u_4 + F \\
\dot{u}_5 &= 3u_2u_4 - 9u_5
\end{aligned}
\end{equation}

Here, $F$ is the external forcing and time is expressed in units of the intrinsic nonlinear timescale of each mode. The system is integrated using an explicit Runge-Kutta-Dormand-Prince method with tolerance $10^{-10}$.

{In the following discussion, the forcing values are set to $F=33$ and to $F=28.718$. These two values were chosen because they identify the {onset of transition to the turbulent regime} ($F=28.718$), where multiple bifurcations begin to open~\cite{Franceschini1979}, and a stage of {``fully developed'' turbulence} ($F=33$).

\begin{figure}[h]
    \centering
    \includegraphics[scale=0.5]{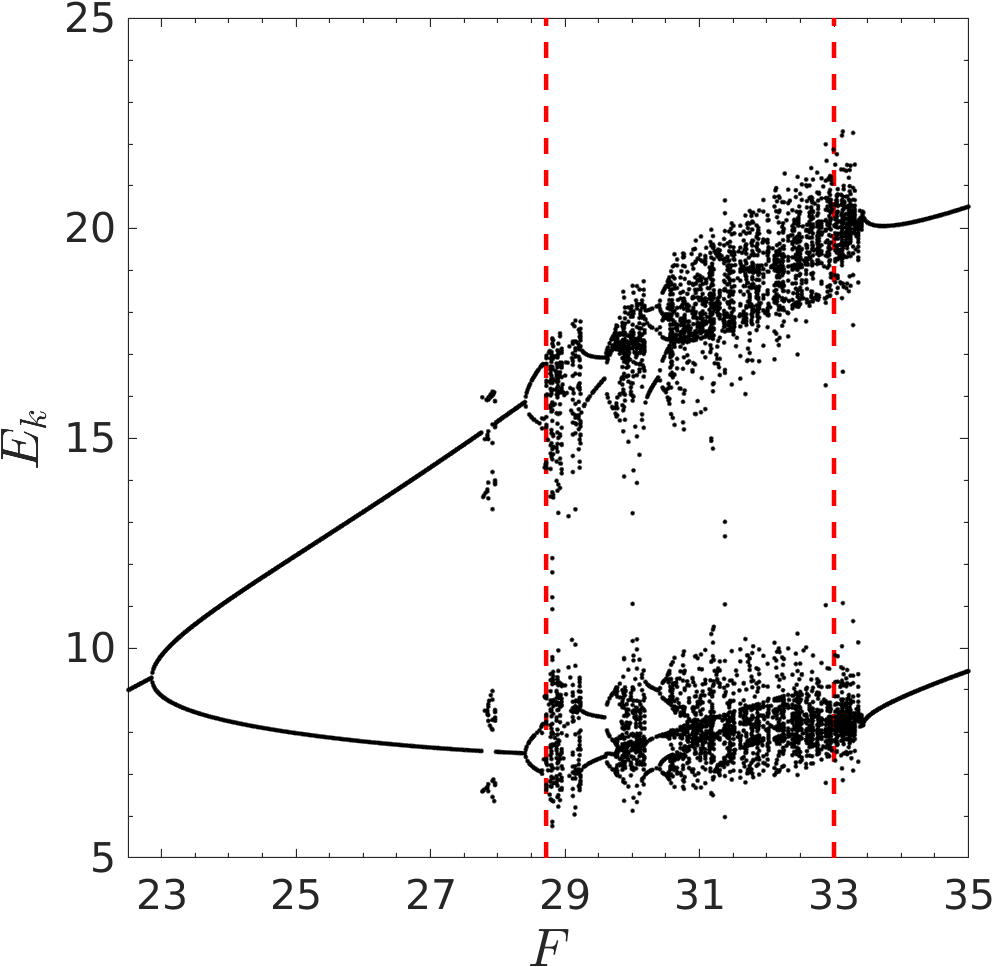}    
   \caption{Bifurcation map obtained from kinetic energy $E_k$ in a range of kinetic Reynolds number $F \in [22, 35]$, spanning the entire dynamical behavior of the 2D truncated Navier-Stokes equations. The red vertical dashed line indicates the forcing values used in the analysis.
   }
   \label{fig:map}
\end{figure}
Figure~\ref{fig:map} reports the bifurcation map obtained from the numerical integration of system of equations~\ref{eq:dynsist}, as a function of the kinetic Reynolds number $F$. For low values of $F$, the system is destabilized by a pitchfork bifurcation, and subsequently, at $F\approx23$, the system is driven into a state of periodic oscillations, passing, this time, through an Hopf bifurcation~\cite{Boldrighini1979}. Finally, as $F$ increases, the system undergoes an infinite sequence of period-doubling Hopf bifurcations, ultimately leading to a fully turbulent regime with a Feigenbaum-type transition to chaos~\cite{Feigenbaum1978}. 
To construct a bifurcation map, the system was integrated over a range of increasing kinetic Reynolds numbers. For each value of $F$, the local maxima and minima of the kinetic energy $E_k$ were recorded once the system reached a steady state or attractor, illustrating the system's response to an external energy injection.

\begin{figure}[h]
    \centering
    \includegraphics[scale=0.26]{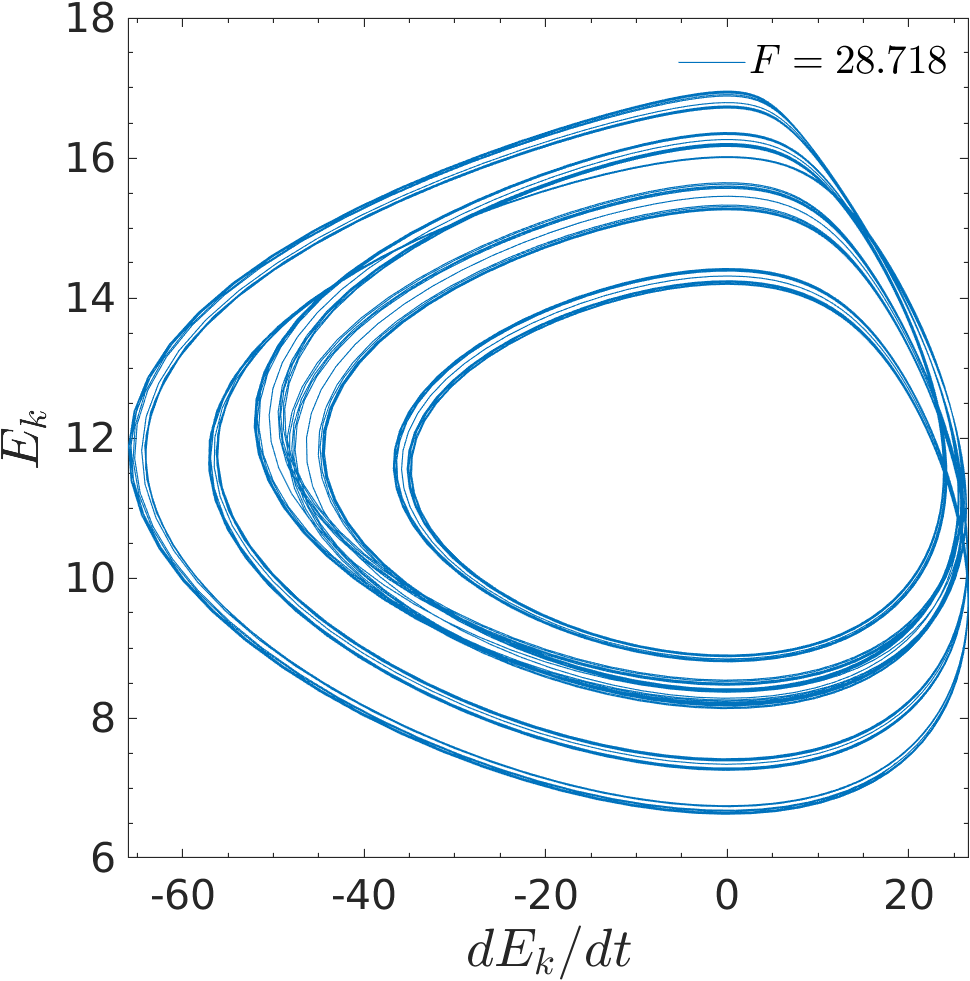}    \includegraphics[scale=0.26]{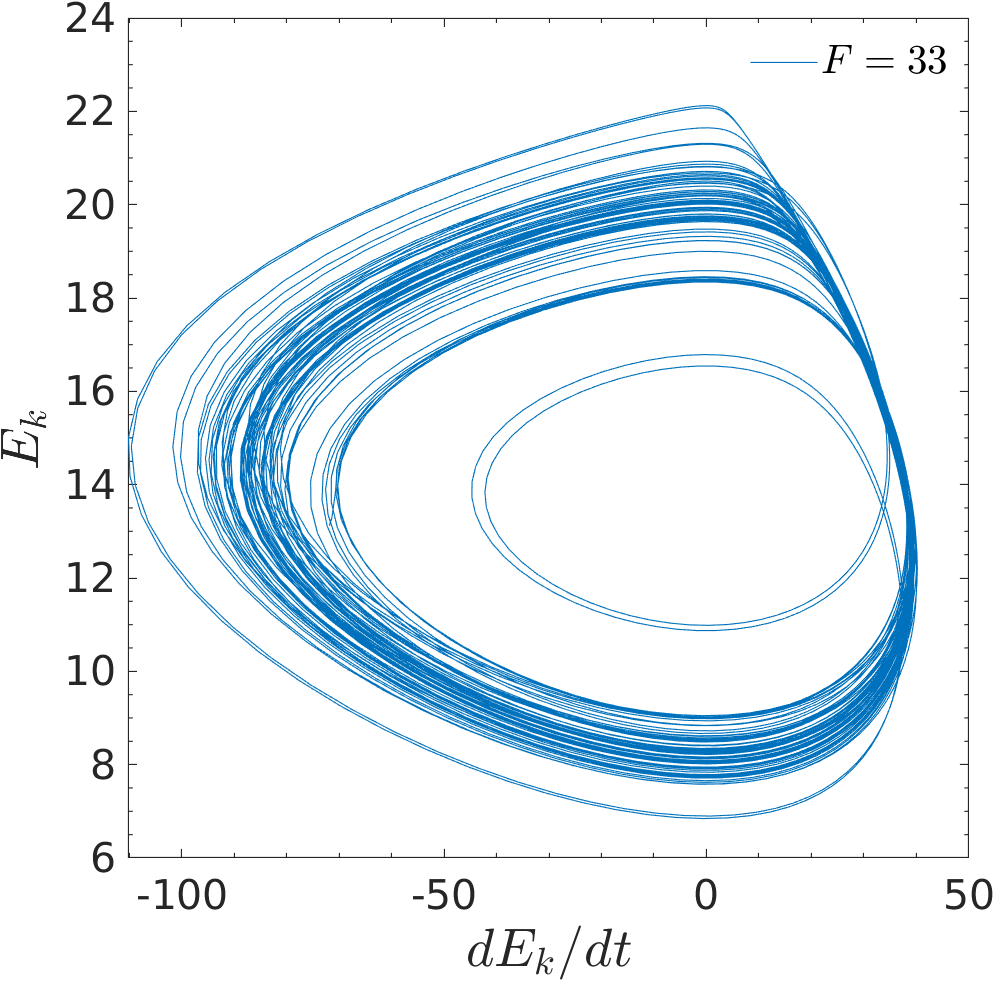}    
   \caption{
   Evolution of the phase space trajectories, projected on the plane $E_k$--$dE_k/dt$, for the two forcing values used in the analysis. Left panel: onset of chaotic region $F=28.718$. Right panel: strange attractor observed for $F=33$, in strong turbulence regime.
   }
   \label{fig:phase_space}
\end{figure}

Figure~\ref{fig:phase_space} reports the evolution of the phase space trajectories, projected onto the $E_k$--$dE_k/dt$ plane, for two different values of $F$. As the forcing is increased from the onset of turbulence $F=28.718$ (left panel of Figure~\ref{fig:phase_space}) to the strongly turbulent case $F=33$ (right panel of Figure~\ref{fig:phase_space}), the dynamics evolve through the opening of multiple quasiperiodic oscillations, characterized by multiple orbits that subsequently close on a periodic solution. This formation of new orbits is characteristic of the period-doubling route to chaos, specifically a Feigenbaum-type transition.

\subsubsection{Lorenz-63}
{The Lorenz-63 model is a simplification of Saltzman's model for Rayleigh–Bénard convection. 
In addition, such model is particularly interesting because it shares a key feature with the Navier-Stokes equations: it exhibits a transition to chaos via a Feigenbaum-type period-doubling cascade~\cite{Franceschini1980}. This shared characteristic makes the Lorenz system a valuable tool for understanding complex dynamics, even in more intricate fluid flows.

Saltzman's original formulation describes thermal convection in a fluid layer heated from below, capturing the complex fluid and temperature dynamics through an infinite set of modes. Lorenz reduced this system by truncating it to only three dominant modes, leading to the well-known three-equation system (Eq. \eqref{eq:lorenz63}).}
 It was introduced by Edward Lorenz in 1963 and is one of the most well-known chaotic systems. The equations are often used as a prototype for studying chaotic dynamics, including phenomena such as sensitivity to initial conditions and deterministic chaos. The system is defined by the following set of three non-linear differential equations:

\begin{equation}    
\begin{aligned}
\label{eq:lorenz63}
\frac{dx}{dt} &= \sigma \left( y - x \right) \\
\frac{dy}{dt} &= x \left( \rho - z \right) - y \\
\frac{dz}{dt} &= x y - \beta z
\end{aligned}
\end{equation}

where \(x\), \(y\), and \(z\) represent the state variables of the system (typically associated with the temperature, the intensity of convection, and the vertical temperature variation, respectively), while \(\sigma\), \(\rho\), and \(\beta\) are system parameters, often referred to as the Prandtl number (\(\sigma\)), the Rayleigh number (\(\rho\)), and the aspect ratio of the convection cell (\(\beta\)). The Lorenz-63 system is known for its chaotic behavior, which arises when the parameters are set to certain values, such as \(\sigma = 10\), \(\rho = 28\), and \(\beta = 8/3\). Under these conditions, the system exhibits a strange attractor, where trajectories in the phase space do not settle into periodic behavior, but instead exhibit sensitive dependence on initial conditions. Small differences in initial conditions lead to vastly different trajectories over time, a hallmark of chaotic systems.

\subsection{Results}\label{sec:NSRes}
To evaluate the performance of our reservoir model, we employ the Valid Prediction Time (VPT)~\cite{VPT_VLACHAS2020191}, defined as the maximum time \(T\) for which the predicted trajectory \(\hat{y}_i(t)
\) remains within an acceptable error range from the corresponding reference trajectory \(y_i(t)\). This can be expressed mathematically as:
\begin{equation}
\text{VPT} = \max \left( T : \forall t \leq T, \, \sqrt{\frac{1}{d} \sum_{i=1}^{d} \left( \frac{\hat{y}_i(t) - y_i(t)}{\sigma_i} \right)^2} \leq \epsilon \right)
\end{equation}
where \(\epsilon = 0.3\) is a predefined error threshold and \(\sigma_i\) is the standard deviation of the time series for each component; the VPT thus quantifies the overall deviation across all dimensions of a vector‐valued time series. 

In the following, we present the results of the application of the algorithm to NS system (Eq. \ref{eq:dynsist1}) and to the Lorenz-63 system (Eq. \ref{eq:lorenz63}).

\subsubsection{Forecasting task and data preparation}
The forecasting task considered in this work is formulated as a discrete-time multivariate prediction problem. Starting from the continuous-time dynamical systems, numerical trajectories are generated through time integration and subsequently sampled at regular intervals $\tau$. The resulting discrete sequence $\mathbf{s}_k = \mathbf{x}(k\tau)$ is used as input to the reservoir, where $\mathbf{x}(k\tau)$ denotes the state vector of the dynamical system sampled at the discrete time step $k\tau$. At each discrete step $k$, the reservoir receives the current system state $\mathbf{s}_k$ and is trained to predict the next state $\mathbf{s}_{k+1}$. Therefore, the model performs one-step-ahead prediction during training. During the testing phase, autonomous forecasting is performed recursively by feeding the predicted output back into the reservoir input. For the Navier–Stokes model, the input vector corresponds to the five-mode amplitudes $\mathbf{s}_k = (u_1,u_2,u_3,u_4,u_5)$, while for the Lorenz-63 system $\mathbf{s}_k = (x,y,z)$. The sampling interval $\tau$ used to discretize the trajectories was set to $\tau=0.015$ for the NS system and to $\tau=0.01$ for the Lorenz-63 system.

\subsubsection{Navier-Stokes}
In a preliminary testing phase, the model was applied without exploiting the time multiplexing. In this case, the vector $\mathbf{m}^k$ of Eq.~\eqref{eq:measvec} consisted of the measurements performed on the initial state $|\psi_0\rangle$ evolved over a single time $\Delta t$. However, when assessing the predictive performance of the algorithm for a range of values of $\Delta t$, it appears that the VPT reaches a peak slightly above 10 time steps, as illustrated in Fig.~\ref{fig:NStimemux} (red markers). This corresponds to a duration significantly shorter than the characteristic time of the NS dynamics.

\begin{figure}[h!btp]
    \centering
    \includegraphics[width=0.9\linewidth]{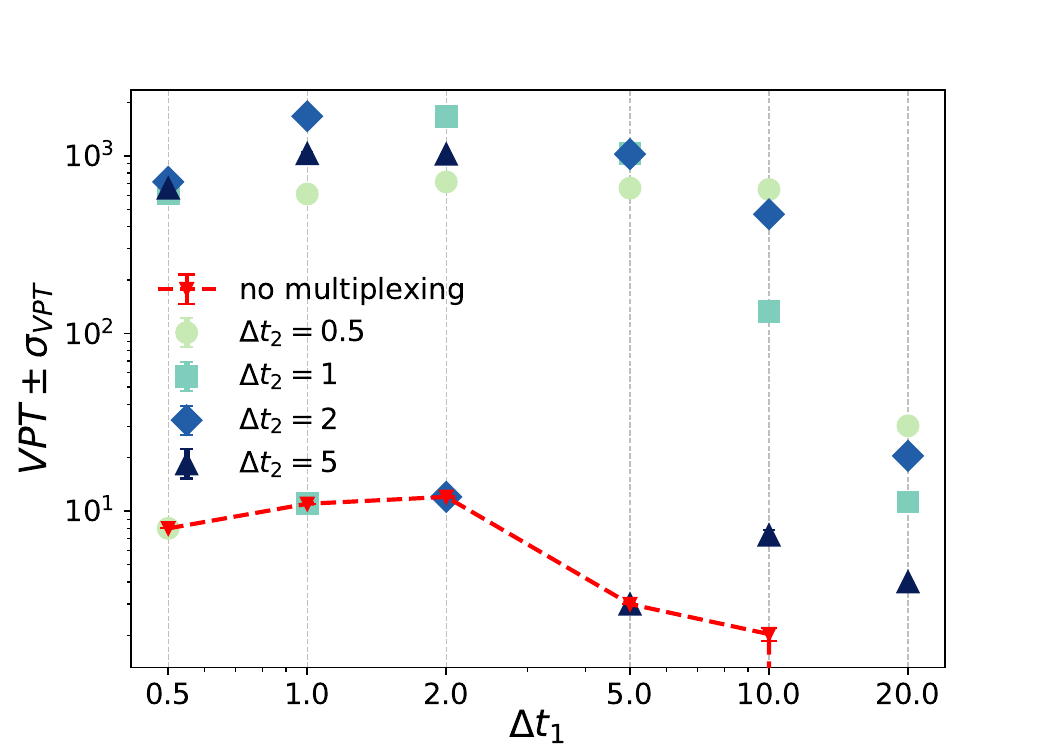}
   \caption{\small{\textbf{Temporal multiplexing}. Comparison of the algorithm performance (VPT) as a function of the quantum system evolution times $\Delta t_1 $, $\Delta t_2.$ Red markers represent the VPT in the absence of time multiplexing (i.e., using a single evolution time $\Delta t_1$). Temporal multiplexing improves the reservoir's ability to capture the system's dynamics by combining intermediate evolution times $\Delta t_1$ and $\Delta t_2$. The coupling \(J\) and the transverse magnetic field \(h\) are set to 0.01 and 0.1, respectively. VPT is reported in discrete time steps.
}}
    \label{fig:NStimemux}
\end{figure}
On the other hand, the temporal multiplexing approach leads to a substantial improvement in prediction performance: the VPT increases by two orders of magnitude for the majority of choices for the evolution times $\Delta t_1, \Delta t_2$. This suggests that with the proposed strategy of concatenating two measurement vectors, each obtained after evolving the state under a distinct time interval, the quantum system is able to extract a richer representation of the underlying NS dynamics. The VPT drops dramatically when \( \Delta t_1 = \Delta t_2 \), which essentially corresponds to the absence of time multiplexing, as reported in Fig.~\ref{fig:NS5_C}. The same figure also clearly shows that the algorithm is invariant under the exchange of \( \Delta t_1\) and \(\Delta t_2 \), as the performance remains unchanged when the two time parameters are swapped.

\begin{figure}[h!btp]
    \centering
    \includegraphics[width=1\linewidth]{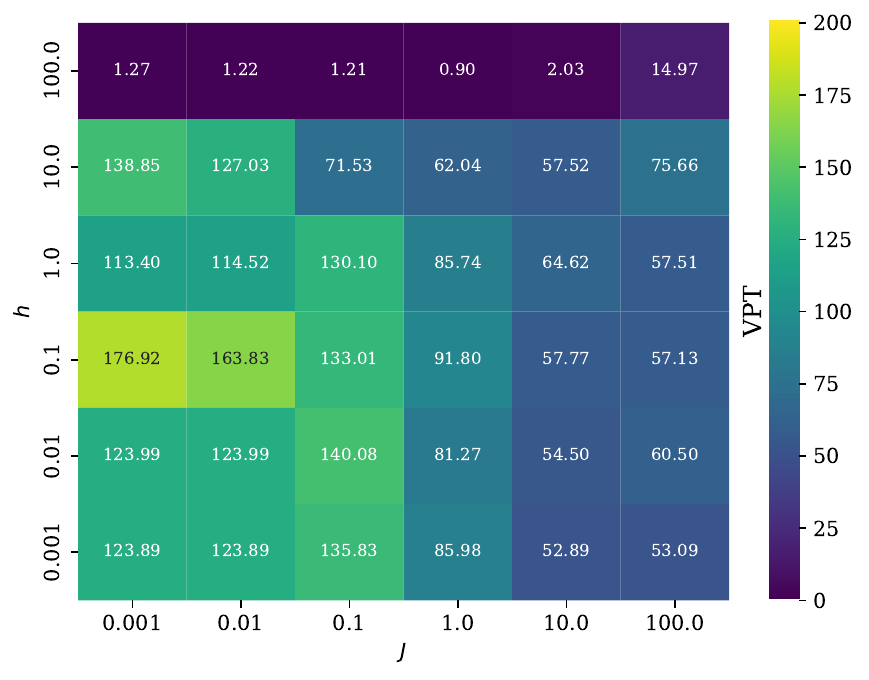}    
   \caption{\small{\textbf{Analysis of parameters}. Variation of the VPT, for NS system with a forcing value set to \(F = 33 \), as a function of the transverse magnetic field $h$ and the coupling $J$ for fixed times $\Delta t_1=1$ and $\Delta t_2=5$. The value of the VPT is normalized with respect to the longest nonlinear time \( T_{k_i} \) among the five oscillation modes of the system and averaged over 30 realizations of $\hat{H}_k$.
}}
   \label{fig:VPT_NSC_JH}
\end{figure}

Further investigating the properties of the quantum system to analyze the performance of our algorithm, we focused on two key parameters of the Hamiltonian $\hat{H}_k$: the transverse magnetic field $h$ and the coupling constant $J$. To achieve this, we selected the pair of intermediate times $\Delta t_1$ and $\Delta t_2$ that maximize the performance for each pair of $h$ and $J$. From the analysis shown in Fig. \ref{fig:VPT_NSC_JH}, it emerges that the best performance is achieved for small values of $J$. It is evident that the performance of the algorithm deteriorates when the transverse magnetic field \( h \) is significantly larger than the input-dependent term \( {h}_k^{i} \). This is because a strong transverse magnetic field reduces the influence of the input-dependent magnetic field on the system's dynamics.
For the same reason, the algorithm's performance also starts to decrease when $J$ becomes comparable in magnitude to the input-dependent magnetic field.

The analysis of the heatmap in Fig. \ref{fig:NS5_C} (Top) reveals the existence of preferential time scales for which the predictive performance is significantly enhanced. As expected, the results of the performances are almost symmetric under exchange of the two time parameters $\Delta t_1$ and $\Delta t_2$. The value of the VPT for $\Delta t_1 = \Delta t_2$ further confirms that a single evolution time is insufficient to fully capture the underlying dynamics of the system. Moreover, there exists a region in the parameter space where the predictive performance deteriorates substantially. This happens when one evolution time is much larger than a certain threshold,  $\Delta t _l \gtrsim 10$ in units of the inverse of the maximum input-dependent magnetic field. In this region, the performance of the algorithm is poorly dependent on the specific values of $\Delta t_l$. 
As shown in Fig. \ref{fig:NS5_C} (Bottom), the relative error
highlights that, in the temporal regions where the VPT is high, it is also robust with respect to different realizations of \(\hat{H}_k\). 
\footnote{The relative error reported throughout the manuscript is defined as the ratio between the standard deviation of the VPT, computed over different realizations of the Hamiltonian parameters, and the corresponding average VPT value. Explicitly, we define $\epsilon_{rel}$=$\sigma_{VPT}/\langle VPT \rangle$ where $\sigma_{VPT}$ and $\langle VPT \rangle$ denote, respectively, the standard deviation of the VPT distribution and its average value.}
   

\begin{figure}[t]
    \centering
    \includegraphics[width=0.9\linewidth]{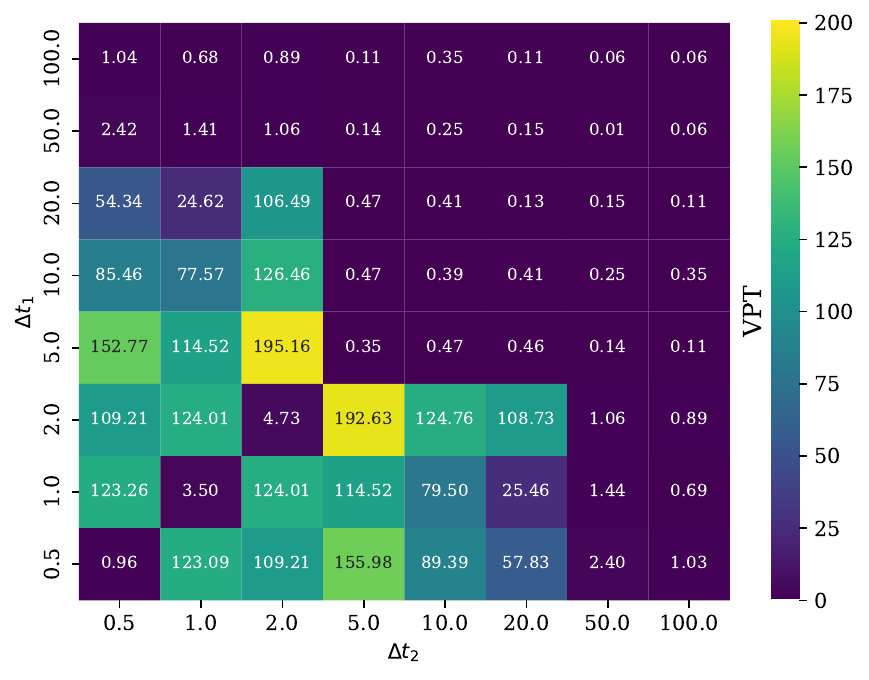}
    \includegraphics[width=0.9\linewidth]{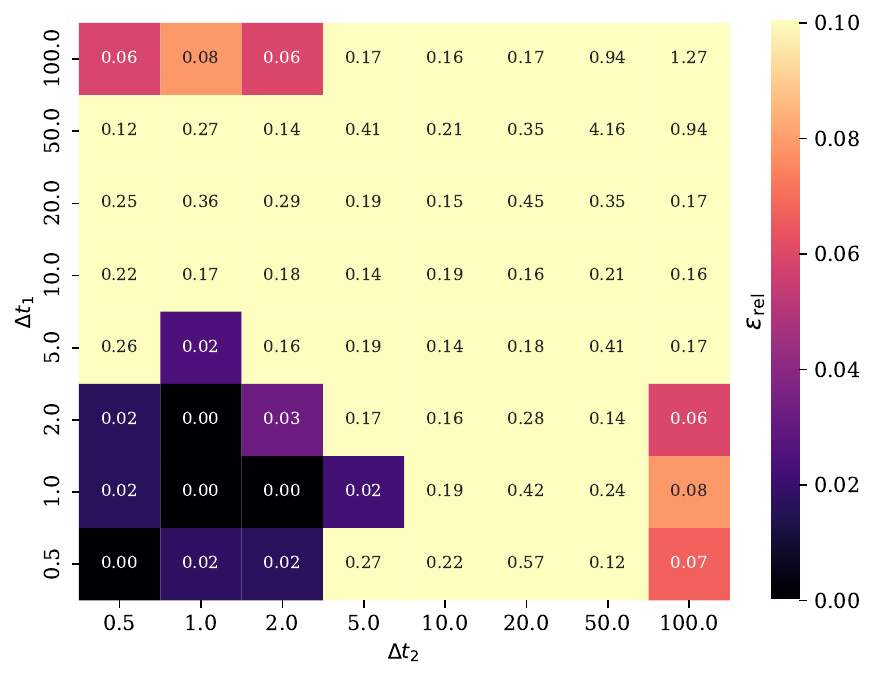}
   \caption{\small{\textbf{Performance of algorithm}. \textbf{Top}: Algorithm performance as a function of the quantum system evolution times $\Delta t_1 $, $\Delta t_2$. The coupling \(J\) and the transverse magnetic field \(h\) are set to $1$ and $0.01$, respectively.} \textbf{Bottom}: Relative error $\epsilon_{rel}$ of the VPT. The coupling J and the transverse field h are set to $1$ and $0.01$, respectively. The results correspond to the NS system with a forcing value set to \(F = 33 \), obtained over 30 realizations of $\hat{H}_k$.}
    \label{fig:NS5_C}
\end{figure}

As shown in Fig. \ref{fig:VPT_NS5_A}, by reducing the dynamical forcing to \( F = 28.718 \), the algorithm's performance improves both in the time domain (Top) and in the parameter space of the coupling strength \( J \) and the transverse field \( h \) (Bottom). 
The relative error of VPT \( \sigma_{\mathrm{VPT}} \) (Fig. \ref{fig:STD_NS5_A}, Top) is negligible precisely in the region of time parameters that maximize the prediction performance of the algorithm. In Fig. \ref{fig:STD_NS5_A} (Bottom), we observe that  \( \sigma_{\mathrm{VPT}} \) remains negligible for small values of \( J \), but increases as \( J \) grows. 
This behavior is explained by the fact that the stochasticity of the Hamiltonian \( \hat{H_k} \) is introduced entirely through the random couplings \( J_{ij} \). Increasing the global coupling parameter \( J \) amplifies the effect of these random fluctuations on the system dynamics, thereby leading to greater variability in prediction performance.  

\begin{figure}[h!btp]
    \centering
    \includegraphics[width=0.9\linewidth]{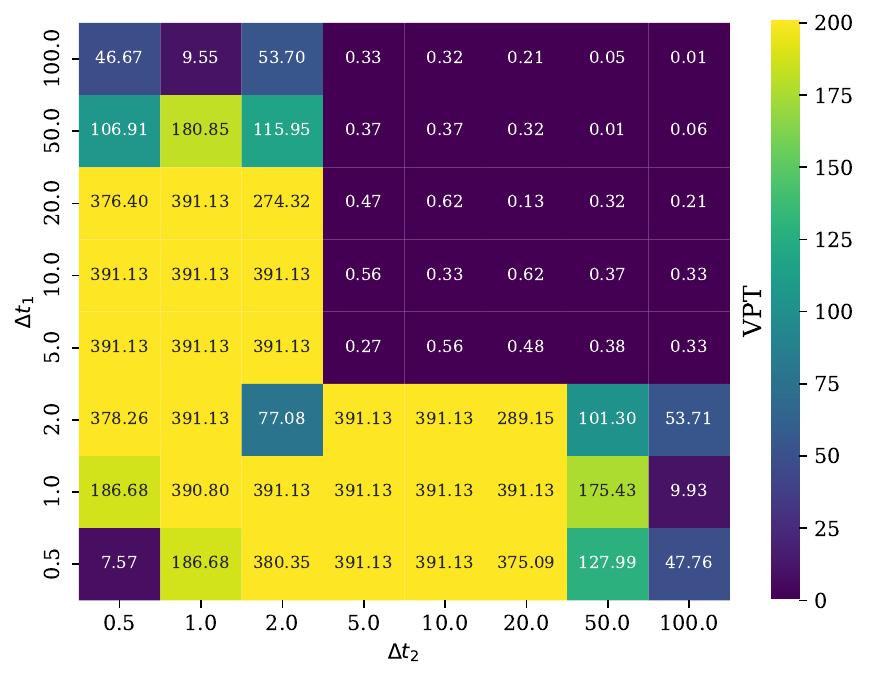}
    \includegraphics[width=0.9\linewidth]{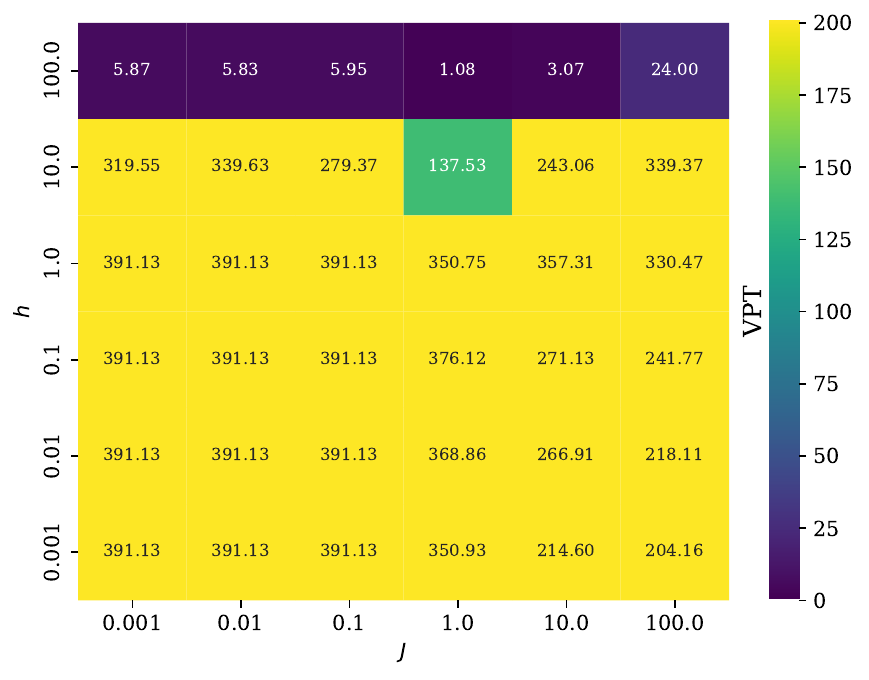}
   \caption{\small{\textbf{Comparison of performance}. Results of the algorithm's performance as a function of the time parameters \(\Delta t_1, \Delta t_2\) \textbf{(Top)} and the spatial parameters \( J \) and \( h \) \textbf{(Bottom)}, obtained over 30 realizations of $\hat{H}_k$. The results correspond to the NS system with a forcing value set to \( F = 28.718 \). The coupling \(J\) and the transverse magnetic field \(h\) are set to $0.01$ and $0.1$ \textbf{(Top)} and  $\Delta t_1=2$, $\Delta t_2=1$\textbf{(Bottom)}.}}
    \label{fig:VPT_NS5_A}
\end{figure}

\begin{figure}[h!btp]
    \centering
    \includegraphics[width=0.9\linewidth]{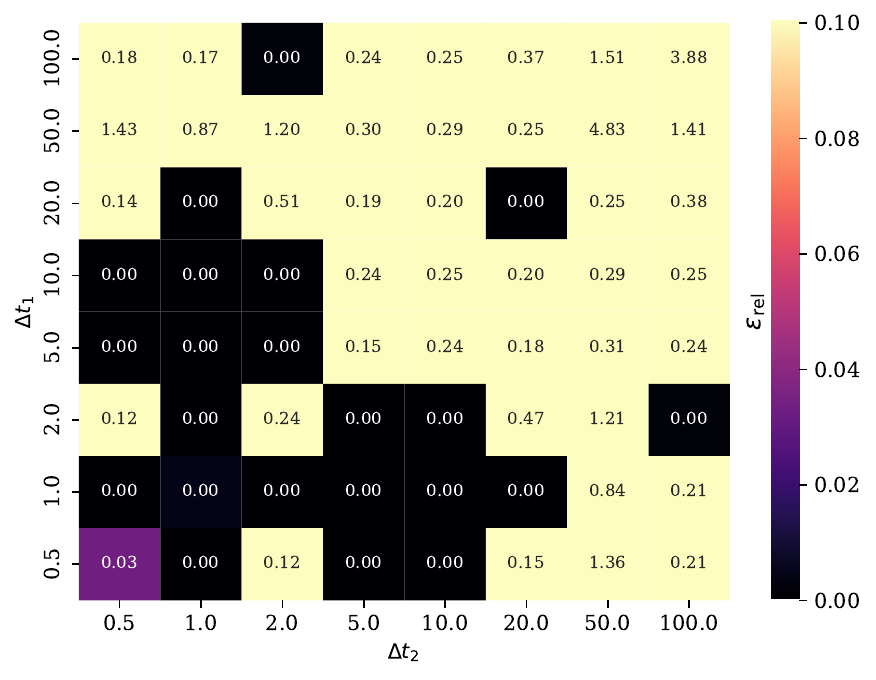}
    \includegraphics[width=0.9\linewidth]{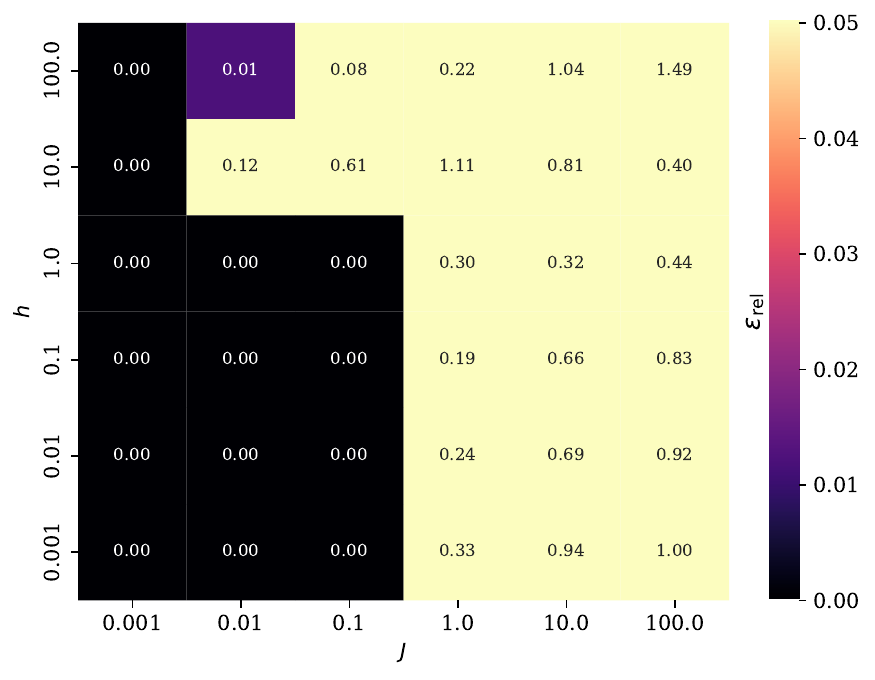}
   \caption{\small{\textbf{Error analysis}. Relative error $\epsilon_{rel}$ of VPT as a function of the time parameters \(\Delta t_1, \Delta t_2\) (Top) and the spatial parameters \( J \) and \( h \) (Bottom). The results correspond to the NS system with a forcing value set to \(F = 28.718 \). The coupling \(J\) and the transverse magnetic field \(h\) are set to $0.01$ and $0.1$ (Top) and  $\Delta t_1=2$, $\Delta t_2=1$(Bottom).}}
     \label{fig:STD_NS5_A}
\end{figure}

\begin{figure}[h!btp]
    \centering
    \includegraphics[width=1\linewidth]{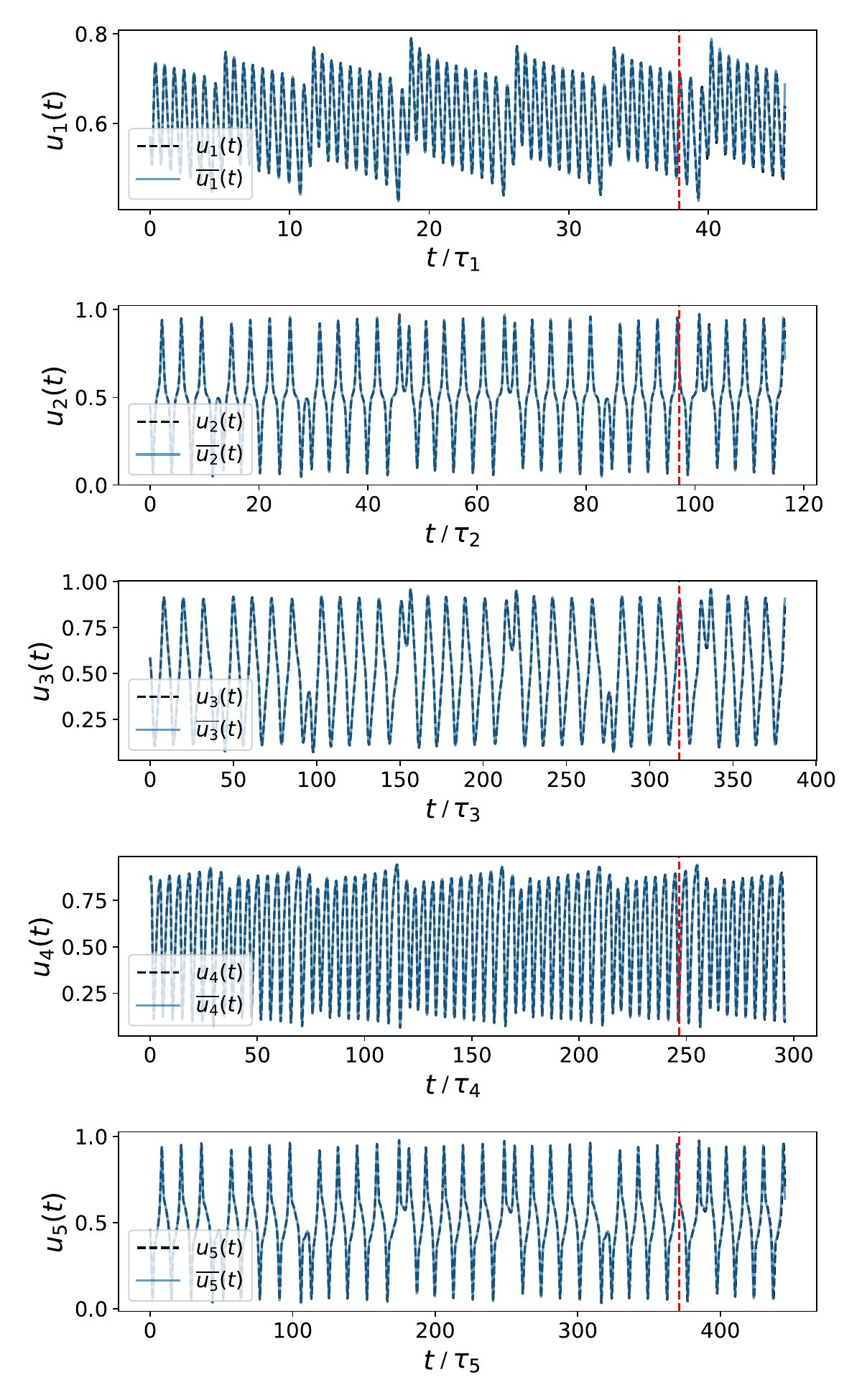}
   \caption{\small{\textbf{Prediction Results.} The figure shows the evolution of the oscillation modes of the input signal of NS time series $(u_{k}(t))$ compared to those predicted by the algorithm $(\overline{u_{k}(t)})$. The time axis is normalized by the nonlinear time of each mode $u_\mathbf{k}$. We have predicted up to approximately 2500 time steps, corresponding to the VPT value indicated by the dashed red line. The length of the training period was set to $N_{tr} = 4500$ time steps.}}\label{fig:NSPred}
\end{figure}

Once the optimal set of parameters has been identified, which maximizes the prediction performance, we present in Fig.~\ref{fig:NSPred} a comparative plot between the original signal numerically generated as a solution of the NS equations truncated via a 5-mode Galerkin projection and a selected sample of the signal obtained using our QRC algorithm. Predictions extend to 2500 time steps, corresponding to a VPT of approximately 38 nonlinear times for the slowest oscillation mode and 370 nonlinear times for the fastest mode.

\paragraph{Noise Analysis}

To assess the robustness of the proposed framework against experimental imperfections, we investigated the impact of measurement noise on the reservoir observables. To this end, at each time step, the expectation values were perturbed according to:
\begin{equation*}
   \langle \hat{O}_i \rangle \rightarrow \langle \hat{O}_i \rangle + \delta_i,
\end{equation*}
where $\delta_i$ is sampled from a uniform distribution in the interval $[-\eta/2,\eta/2]$, with $\eta$ controlling the noise amplitude. This simple additive noise model is intended to mimic generic measurement fluctuations that may occur in practical quantum hardware implementations.
Figure~\ref{FigsNew:Noise} reports the resulting VPT as a function of the noise strength $\eta$. As the noise amplitude increases, the prediction accuracy progressively deteriorates, resulting in a reduction of the VPT.
Overall, these results indicate that the proposed hybrid quantum reservoir computing architecture exhibits a significant degree of robustness against moderate measurement noise, maintaining meaningful forecasting capabilities in the low-noise regime. The additive noise model adopted here provides a phenomenological description of uncertainties affecting the measured expectation values. In practical quantum hardware, a major source of such uncertainty originates from the finite number of measurement shots used to estimate expectation values. In this case, the statistical error decreases as $1/\sqrt{N_{\mathrm{shots}}}$, where $N_{\mathrm{shots}}$ denotes the number of repeated measurements \cite{Gebhart2023}. Therefore, the noise amplitude $\eta$ considered in our simulations can be interpreted as an effective measure of the uncertainty associated with finite sampling.

\begin{figure}[h!btp]
    \centering
    \includegraphics[width=0.9\linewidth]{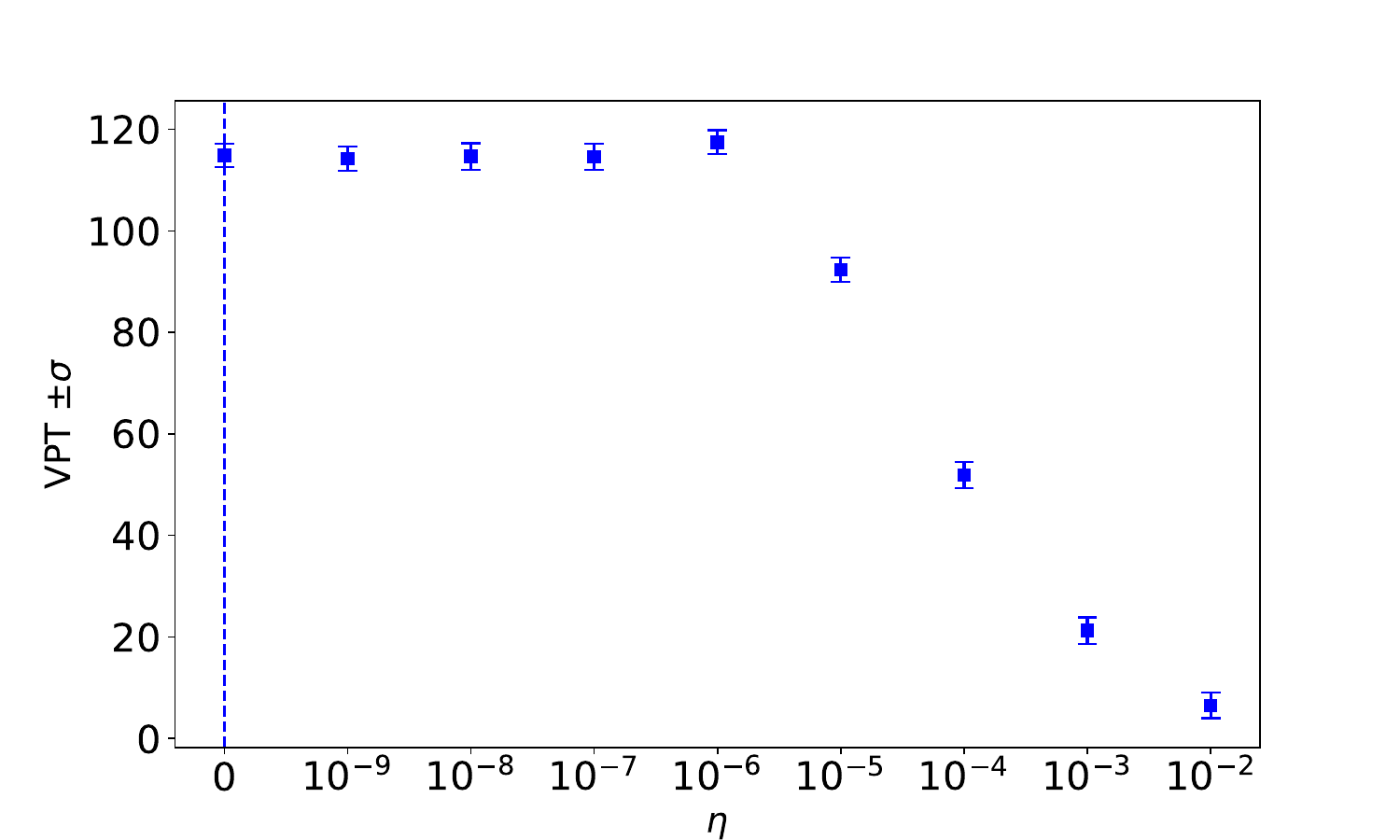}
   \caption{\small{\textbf{Noise Analysis}. VPT for the NS series prediction as a function of the noise strength $\eta$. For each point, the mean value and the standard deviation
of the mean are shown, based on 100 different realizations of $\hat{H}_k$ and random noise. The vertical dashed line indicates
zero-noise simulation. The value of the VPT is normalized with respect to the longest nonlinear time \( T_{k_i} \) with $F=33$.}}
    \label{FigsNew:Noise}
\end{figure}

\subsubsection{Lorenz-63}
In order to validate the effectiveness of our model against established benchmarks in the scientific literature, we apply the proposed hybrid QRC model to the Lorenz-63 \cite{Lorenz1963} system, a widely studied prototype for deterministic chaos and nonlinear dynamics.
The dynamics of chaotic systems can be quantitatively characterized by the leading Lyapunov exponent \(\lambda_L\), which quantifies the average exponential divergence of initially close trajectories. 
This exponent sets a fundamental time scale for chaos, determining the temporal horizon beyond which accurate predictions become unreliable. Consequently, we have rescaled the VPT in units of \emph{Lyapunov time} (LT), defined as ${1/\lambda_L}$.
\begin{figure}[h!btp]
    \centering
    \includegraphics[width=0.9\linewidth]{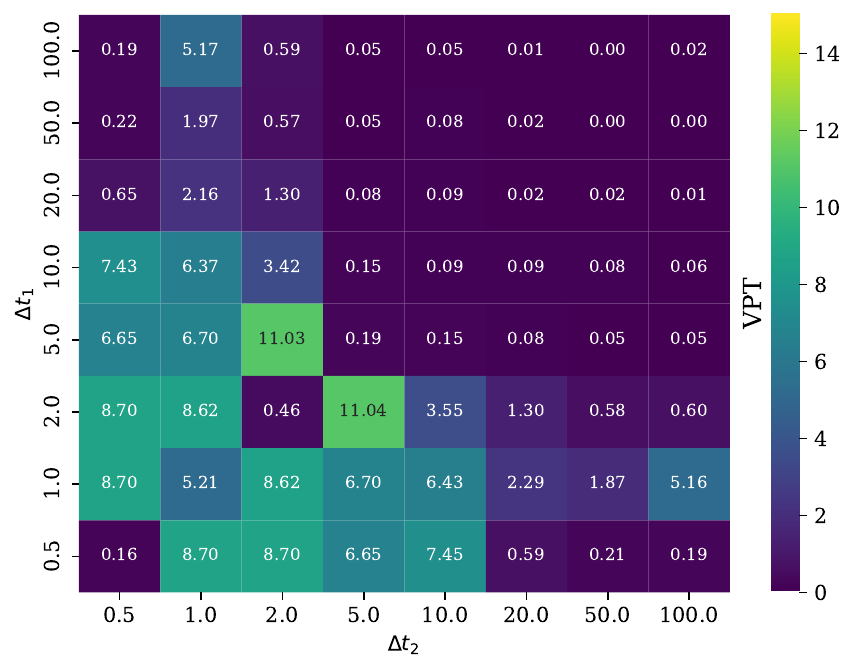}
    \includegraphics[width=0.9\linewidth]{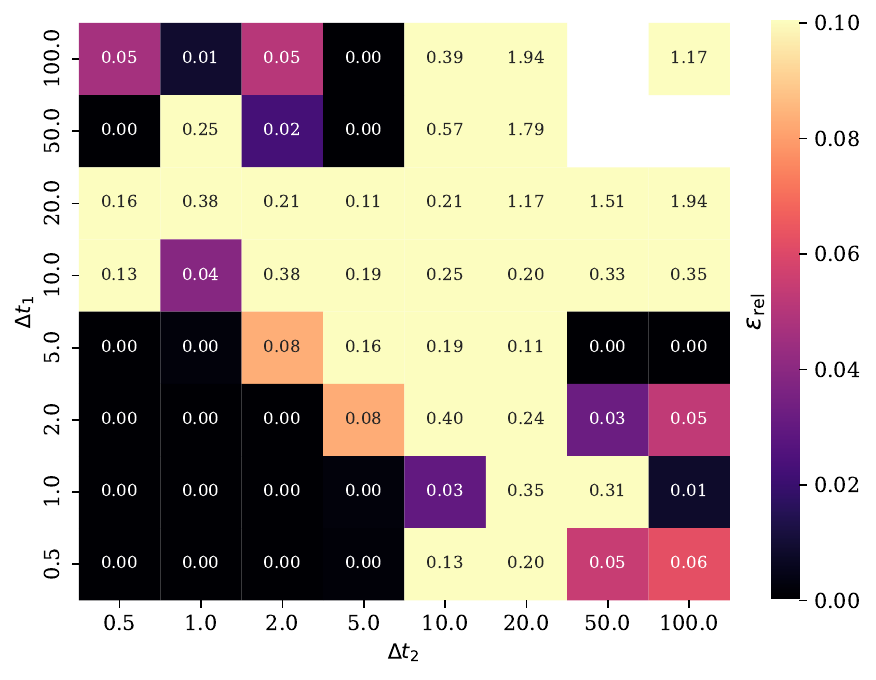}
   \caption{\small{\textbf{Performance}. \textbf{Top}: Average value of the VPT (expressed in LT), computed over 30 realizations of $\hat{H}_k$. Predictions were carried out up to 500 time steps, corresponding to about 13 LT.} \textbf{Bottom}: Relative error $\epsilon_{rel}$ of the VPT.  The coupling \(J\) and the transverse field \(h\) are set to 0.001 and 0.001, respectively. Cells with missing numerical values correspond to realizations of $\hat{H}_k$ for which VPT is zero.}
    \label{fig:VPT_STD_L63}    
\end{figure}
Compared to the performance observed with the Navier-Stokes model, the Lorenz-63 system exhibits a significantly narrower region of optimal algorithm functioning with respect to the evolution times \(\Delta t_1\) and \(\Delta t_2\). As shown in Fig. \ref{fig:VPT_STD_L63} (Top), for the pair \((J, h)\) that maximizes performance, there is one pair of evolution times that allows for sufficiently long prediction horizons relative to the intrinsic timescales of the system. Comparing Fig.~\ref{fig:VPT_STD_JH_L63} (Top) with Figs. \ref{fig:VPT_NSC_JH} and \ref{fig:VPT_NS5_A} (Bottom), we can observe that the dynamics of the quantum reservoir that allow a long-term prediction is independent on the specific time series.  In other words, the optimal region in the parameter space \( (J, h) \) for the Lorenz-63 prediction task corresponds to the one for the Navier-Stokes time series. The same considerations made for the Navier-Stokes system apply to the relative error analysis shown in Figs.~\ref{fig:VPT_STD_L63} and~\ref{fig:VPT_STD_JH_L63} (Bottom).

\begin{figure}[h!btp]
    \centering
    \includegraphics[width=0.9\linewidth]{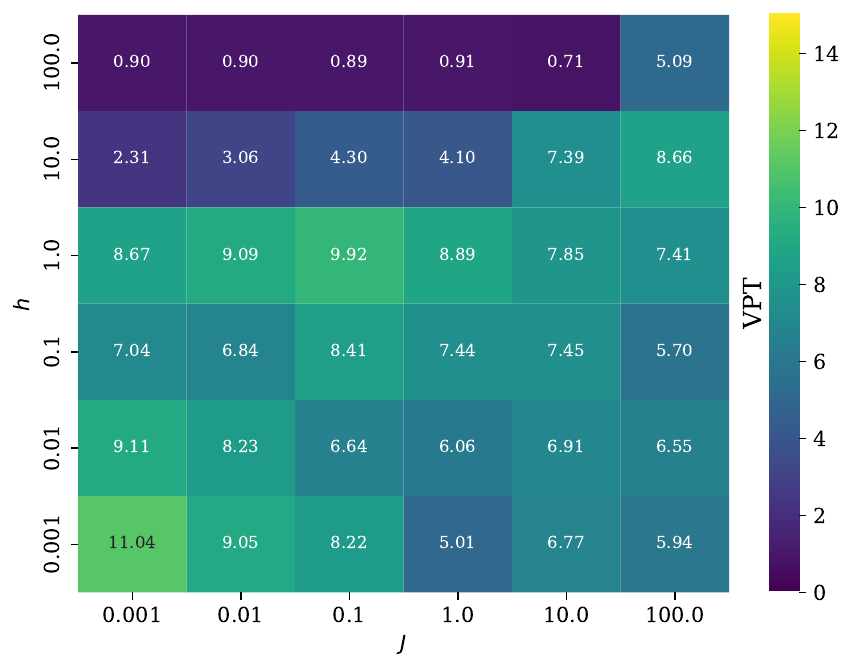}
    \includegraphics[width=0.9\linewidth]{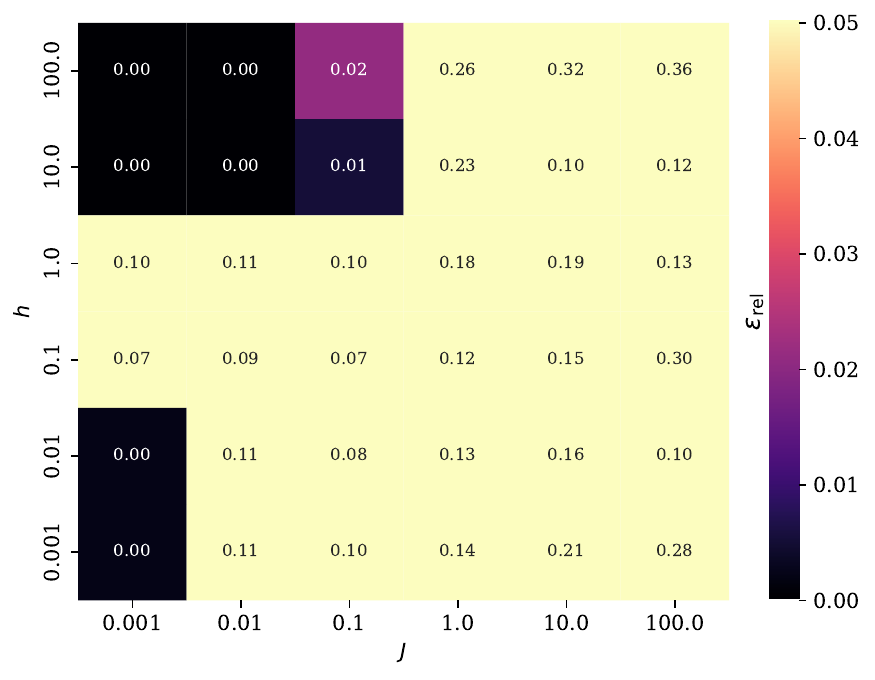}
    \caption{\small{\textbf{Analysis of parameters}. \textbf{Top}: Variation of the VPT (expressed in LT) as a function of the transverse magnetic field $h$ and the coupling $J$ for fixed times $\Delta t_1=2$ and $\Delta t_2=5$, including the relative error $\epsilon_{rel}$ (\textbf{Bottom}), obtained over 30 realizations of $\hat{H}_k$ .}}
      \label{fig:VPT_STD_JH_L63}
\end{figure}

\begin{figure}[h!tbp]
    \centering
    \includegraphics[width=0.9\linewidth]{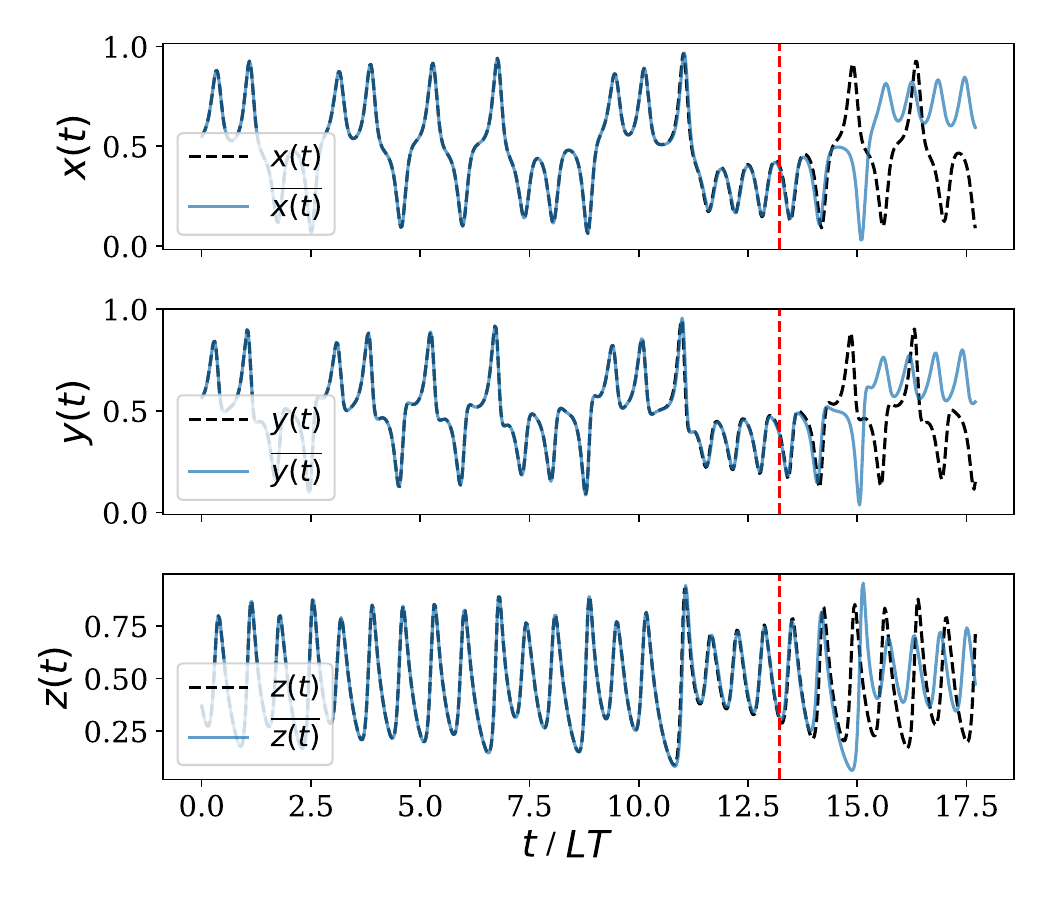}
 
   \caption{\small{\textbf{Prediction Results}. Evolution of the Lorenz-63 system $(x(t),y(t),z(t))$, compared with the signals predicted by the QRC algorithm $(\overline{x(t)},\overline{y(t)},\overline{z(t)})$. The best prediction extends up to approximately 500 time steps, corresponding to about 13 LT, indicated by the dashed red line. The length of the training period was set to $N_{tr} = 1300$ time steps.} }
        \label{fig:L63_PRED}
\end{figure}
Our QRC algorithm is capable of providing relatively long-term predictions (Figs. \ref{fig:VPT_STD_JH_L63} and \ref{fig:L63_PRED}) compared to those reported in the literature \cite{wudarski2024arxiv,Ahmed2024Prediction}. It can be observed that the \(z\)-component of the signal allows for more robust predictions than the other two components. 

\section{Conclusions}\label{sec:conclusions}

In this work, we have developed and tested a hybrid quantum reservoir computing algorithm for multidimensional data that integrates quantum dynamical evolution with classical memory enhancement. The model has been evaluated on two representative chaotic systems: a five-mode Galerkin truncation of the Navier-Stokes equations and the Lorenz-63 system.

Our results show that the QRC algorithm is capable of capturing complex nonlinear dynamics and achieving competitive prediction performance in terms of the Valid Prediction Time, provided one uses temporal multiplexing with (at least) two different time evolutions of the quantum system. Our analysis reveals that the algorithm performs optimally for specific ranges of the evolution times $\Delta t_1$ and $\Delta t_2$, and when the Hamiltonian parameters of the quantum system — coupling $J$ and transverse field $h$ — are appropriately tuned with respect to the scale of the encoded input signal. Furthermore, the observed consistency in optimal parameter regions across different dynamical regimes and for both chaotic systems we investigated points to a form of robustness of the quantum encoding scheme, which could prove valuable for a wider class of dynamical systems without extensive re-optimization.

Overall, these findings indicate that hybrid quantum reservoir computing may provide an effective framework for modeling complex nonlinear dynamics using relatively small quantum systems. In the present approach, the role of the quantum reservoir is not simply tied to the asymptotic growth of Hilbert-space dimension, but rather to its ability to generate rich nonlinear dynamical feature maps through coherent many-body evolution and multi-observable measurements. The classical memory layer complements the quantum dynamics by enhancing temporal information retention.

Beyond the specific forecasting tasks considered here, our results suggest a more general perspective on hybrid quantum reservoir computing. The quantum dynamical evolution acts as a high-dimensional nonlinear feature generator whose characteristic time scales determine the effective representation of the input dynamics. When these intrinsic time scales become commensurate with those governing the target system, the reservoir naturally develops a representation capable of capturing the underlying chaotic attractor. From this viewpoint, the predictive capability of the hybrid architecture may not be entirely tied to a specific model, but rather on a dynamical resonance between the quantum evolution and the characteristic scales of the input process. This observation hints at a possible universality in the design of quantum reservoirs for nonlinear dynamical systems, where suitable regimes of the quantum dynamics may provide generic embeddings for classes of chaotic signals.

An additional relevant aspect of the proposed architecture is its compatibility with near-term quantum hardware. Since the method relies on fixed quantum dynamics and does not require the variational optimization of quantum gates, the quantum subsystem can operate as a passive nonlinear feature generator. Moreover, the use of a small number of qubits together with the classical memory layer substantially reduces the quantum hardware requirements. The protocol could in principle be implemented on several current quantum computing platforms, including superconducting circuits, trapped ions, neutral atoms, and photonic processors, where measurements of local observables and short-time coherent dynamics are experimentally accessible. In this context, temporal multiplexing may provide an effective strategy to enhance the expressive power of small-scale quantum devices without increasing the physical system size.

Future work will focus on scaling the approach to higher-dimensional reservoirs and exploring alternative quantum architectures to enhance expressivity and resilience. Additionally, extending the framework to include training of quantum parameters, or adaptive control of temporal multiplexing strategies, could further improve prediction capabilities and generalization. 
Moreover, a deeper theoretical understanding of the interplay between quantum dynamics, temporal multiplexing, and memory capacity would further strengthen the foundation of quantum reservoir computing as a viable tool in the study and prediction of non-linear phenomena.

\section{Acknowledgements}
We thank N. Lo Gullo for fruitful discussions.
We acknowledge the use of the Alarico high-performance computing cluster provided by the University of Calabria.

\section{Funding}
	F.C., J.S., C.G. and L.P. acknowledge the contribution received from program Fondo per il
	Programma Nazionale di Ricerca e Progetti di Rilevante Interesse Nazionale
	(PRIN) under the project ``TURBIMECs: study of TURBulence In MEditerranean Cyclone events'', grant n. 2022S3RSCT, CUP Master B53D2300750000.

    This work has been  supported by project FAIR - Future AI Research - Spoke 9 (Directorial Decree no. 1243, August 2nd, 2022; PE 0000013; CUP B53C22003630006), under the NRRP (National Recovery and Resilience Plan) MUR program (Mission 4, Component 2 Investment 1.3) funded by the European Union – NextGenerationEU, and by the PNRR MUR project PE0000023-NQSTI through the cascaded projects “QuCADD” and “ThAnQ”,

\section{Author contributions}

L. S.: Data curation (lead), Formal analysis (lead), Investigation (lead), Visualization (lead), Writing – original draft (lead)
L. M.: Software (lead), Data curation (lead), Investigation (lead), Visualization (lead), Writing – original draft (support), Writing – review (support)
A. G.: Conceptualization (support), Methodology (support), Software (support), Writing – review (support) 
F. D.: Software (support), Writing – review (support)
C. M.: Writing – review (support), Supervision (support), Funding acquisition (support), Resources (support)
L. P.: Writing – review (support), Funding acquisition (support)
A. V.: Writing – review (support), Resources (support)
C. G.: Writing – review (support), Funding acquisition (support)
L. P.: Methodology (support), Writing – review (support), Funding acquisition (support)
F. P.: Conceptualization (support),  Writing – review (support), Supervision (lead), Funding acquisition (support)
J. S.: Conceptualization (lead), Methodology(lead), Data curation (support), Visualization(support), Supervision (lead), Funding acquisition (support), Writing – original draft (support), Writing – review (support)
F. C.: Conceptualization (support), Funding acquisition (support), Project administration (lead). 
Corresponding author: J. S. (jacopo.settino@unical.it).

\section{Competing interests}
The authors declare no competing interests.
\clearpage

\newpage
\pagebreak 
\appendix

\section{A Low Dimensional Truncation of the 2D Navier-Stokes Equations}
\label{app:NS5}

The non-dimensional Navier-Stokes (NS) system describes the dynamics of incompressible fluids within a two-dimensional periodic domain, defined by $0\leq (x,y) \leq L\equiv 2\pi$. It is based on the projections of the momentum equations (Eq.~\ref{eq:mom}) onto the coordinate axes and the continuity equation (Eq.~\ref{eq:cont}).

Given that in fully developed 2D turbulence energy is not only dissipated by viscosity but rather cascades inversely to larger scales~\cite{Boffetta2011}, a Rayleigh-type friction term ($-\lambda_R \mathbf{u}$) is usually introduced into the NS system. This term acts as a sink, dissipating large-scale energy and effectively representing physical processes that the model does not explicitly resolve, such as friction at the bottom of a fluid layer or the aggregated influence of sub-grid scale features.

Within this framework, the NS equations are expressed as:
\begin{align}
\frac{\partial \mathbf{u}}{\partial t} +(\mathbf{u}\cdot\nabla)\mathbf{u} &= -\nabla P + \textnormal{Re}^{-1}\nabla^2\mathbf{u} -\lambda_R \mathbf{u} + \mathbf{F}\label{eq:mom}\\
\nabla\cdot\mathbf{u} &= 0 \label{eq:cont},
\end{align}
where $\mathbf{u}$ represents the velocity vector, $P$ is the kinetic pressure normalized by the fluid density $\rho_0$, $\textnormal{Re}$ is the Reynolds number (incorporating kinematic viscosity), and $\mathbf{F}$ denotes an external forcing term. In 2D, the fields possess only components in the plane: $\mathbf{u}(\mathbf{r},t) = [u_x(x,y,t); u_y(x,y,t)]$ and $P(\mathbf{r},t)=p(\mathbf{r},t)/\rho_0$.

In wave-vector space, the velocity field $\mathbf{u}(\mathbf{r},t)$ is expanded in terms of Fourier coefficients as $\sum_k \mathbf{u}(\mathbf{k},t)e^{-i\mathbf{k}\cdot\mathbf{r}}$, where $\mathbf{k} = 2\pi\mathbf{n}/L$ with $\mathbf{n}\in\mathbb{Z}^2$ being a pair of integers. Due to the divergenceless nature of the fields, the Fourier coefficients $\mathbf{u}(\mathbf{k},t)$ can be defined through a unit polarization vector $\mathbf{e}(\mathbf{k})$ which is perpendicular to the wave-vector (i.e., $\mathbf{k}\cdot\mathbf{e}(\mathbf{k})=0$). Thus, $\mathbf{u}(\mathbf{k},t) = {u}_\mathbf{k}(t)\mathbf{e}(\mathbf{k})$. This unit vector satisfies $\mathbf{e}(\mathbf{k})=\mathbf{e}^\star(-\mathbf{k})$ and $\mathbf{e}(\mathbf{k})\cdot\mathbf{e}^\star(\mathbf{k})=1$, and can be explicitly written as $\mathbf{e}(\mathbf{k}) = i k^{-1}(k_y,-k_x)$, where $k_x$ and $k_y$ are the components of $\bf{k}$ in the plane and $k= |\bf{k}|$.

When projected onto Fourier space, the NS equations transform into an infinite set of ordinary differential equations for the complex amplitudes $u_\mathbf{k}(t)$, which evolve in a $2 \times N$ dimensional space~\cite{Boldrighini1979,Franceschini1984,Carbone2021,Carbone2022c}:
\begin{align}
\label{eq:ODEu}
\frac{d{u}_\mathbf{k}(t)}{d t} &= \frac{4\pi^2}{L^2}\sum_{\mathbf{p},\mathbf{q}}\delta_{\mathbf{k},\mathbf{p}+\mathbf{q}}\;
C_{\mathbf{k}\mathbf{p}\mathbf{q}}[u_\mathbf{p}(t)u_\mathbf{q}(t)] \nonumber  \\
&- \left[\textnormal{Re}^{-1} k^2 + \lambda_R\right] u_\mathbf{k}(t) + F_\mathbf{k}.
\end{align}

where $C_{\mathbf{k}\mathbf{p}\mathbf{q}} = 1/2 [M_{\mathbf{k}\mathbf{p}\mathbf{q}}\pm M_{\mathbf{k}\mathbf{q}\mathbf{p}}]$ are the coupling coefficients of nonlinear terms, with $M_{\mathbf{k}\mathbf{p}\mathbf{q}} = [-i\mathbf{k}\cdot\mathbf{e}(\mathbf{q})][\mathbf{e}^\star(\mathbf{p})\cdot\mathbf{e}(\mathbf{q})]$.
Finally, the external forcing term $F_\mathbf{k}$ (deterministic and constant in time) is introduced, which eventually acts on the system, and the sum in the nonlinear term 
$\sum_{\mathbf{p},\mathbf{q}} \delta_{\mathbf{k},\mathbf{p}+\mathbf{q}}$ is extended to all triads of wave-vectors satisfying the triangular condition $\mathbf{k} = \mathbf{p} + \mathbf{q}$.

In the absence of forcing, dissipation, and drag, the number of wavevectors involved in the nonlinear couplings is infinite. In this inviscid and unforced limit, the system possesses two rugged invariants that survive each single triad of interacting wavevectors~\cite{Moffatt1983,Ting1986,Bartolo2006}: the total kinetic energy $E_k = {1}/{2} \int_A |\mathbf{u}|^2 \, dA$ and the enstrophy $\Omega = {1}/{2} \int_A \omega^2 \, dA$, where $A=[0,2\pi]\times[0,2\pi]$ is the computational domain.

As these rugged invariants are preserved under any Galerkin truncation of the infinite system (Eq.~\ref{eq:ODEu}), a finite Lorenz-like low-order model $\mathbb{L}_N(u)$ can be derived. This model effectively retains all global characteristics of the complete system by considering only a finite sequence of $N$ interacting modes $\mathbf{k}_n$ ($n = 1, 2, \dots,N$). These modes must satisfy the triangular condition $\mathbf{k}_n = \mathbf{k}_{n + r} \pm \mathbf{k}_{n + s}$, where $|n+r| \leq N$ and $|n+s| \leq N$, with $(r, s) \in \mathbb{Z}$.

Furthermore, an analysis of the structure of equations (Eq.~\ref{eq:ODEu}) reveals that the dynamics of a truncated low-dimensional system $\mathbb{L}_N(u)$, regardless of its order $N$, can be decomposed into two subsystems: $\mathbb{L}_N(u)=\mathbb{R}_N(u)\cup\mathbb{I}_N(u)$. This property is particularly significant as it allows the entire system to be reduced to a purely real system, where $\mathbb{I}_N(u) = \emptyset$. Such a property defines an invariant subspace for the system's dynamics, meaning that the system evolves exclusively within the real subspace $\mathbb{R}_N(u)$. This effectively reduces the dimensionality of the system to an $1\times N$ dimensional space.

Here, a five-mode truncated model, $\mathbb{L}_5(u)$ (in a purely real subspace), has been numerically investigated. The selected wave-vectors are $\mathbf{k}_1 = (0, 1)$, $\mathbf{k}_2 = (1, 1)$, $\mathbf{k}_3 = (1, 2)$, $\mathbf{k}_4 = (2, -1)$, and $\mathbf{k}_5 = (3, 0)$, which satisfy triangular relations such as $\mathbf{\mathbf{k}}_1 = \mathbf{k}_3 - \mathbf{k}_2$ and $\mathbf{k}_2 = \mathbf{k}_5 - \mathbf{k}_4$. By exploiting the complex conjugate condition $u_{-\mathbf{k}}(t) = u_\mathbf{k}^\star(t)$ (introduced above) and by explicitly expanding the sums in the nonlinear terms, the dimensionless NS model is reduced to a 5-dimensional autonomous dynamical system. 
This system describes the dynamics of two interacting triads, which contain both local and non-local interactions~\cite{Carbone2024}. In particular, the forcing was applied in a way that ensures an efficient transfer of energy between various modes, favoring the development of the classical bi-directional cascade~\cite{Boffetta2010}.

Upon explicitly calculating the coupling coefficients $C_{\mathbf{k}\mathbf{p}\mathbf{q}}$ and for simplicity setting $\textnormal{Re} = 1$, $\lambda_R = 0$, the truncated NS model can be written as:

\begin{equation}  
\begin{aligned}
\label{eq:dynsist}
\dot{u}_1 &= 3u_3 u^\star_2 -u_1 \\
\dot{u}_2 &= -4u_3u^\star_1 + 4u_5u^\star_5 - 2 u_2 \\
\dot{u}_3 &= u_1u_2 - 5u_3 \\
\dot{u}_4 &= 7u_5u^\star_2 - 5u_4 + F\\
\dot{u}_5 &= 3u_2u_4 - 9u_5 
\end{aligned}
\end{equation}
Here, dotted variables represent time derivatives, {$u^\star_k = u_k$, due to the reality condition $u_k\in\mathbb{R}_N$},
and $F$, hereafter referred to as the kinetic Reynolds number, is the forcing. The condition $F\in \mathbb{R}$ is applied to ensure that the dynamics remain in the real subspace.
Throughout this article, time is given in units of the intrinsic nonlinear timescale $\tau_k$, which characterizes each oscillation mode of the truncated NS system.

The autonomous system was numerically integrated using an explicit Runge-Kutta-Dormand-Prince method~\cite{Dormand1980,Hairer1993}, with an error tolerance set at $10^{-10}$. The invariants of the inviscid and unforced system were conserved within this specified error tolerance. 

\clearpage
\bibliography{ReferencesNEW.bib}

\clearpage

\end{document}